\documentclass[aps,pra,amsmath,amssymb,showpacs,%
floatfix,superscriptaddress]{revtex4}
\usepackage{braket}
\usepackage{graphicx}
\usepackage{dcolumn}
\usepackage{epstopdf}

\newcommand{\Op}[1]{\boldsymbol{\mathsf{\hat{#1}}}}

\def\openone{\leavevmode\hbox{\small1\kern-3.3pt\normalsize1}}

\begin{document}

\title{Optimal control under spectral constraints:\\ Enforcing
  multi-photon absorption pathways}

\author{Daniel M. Reich}
\affiliation{Theoretische Physik,  Universit\"at Kassel,
  Heinrich-Plett-Str. 40, 34132 Kassel, Germany}

\author{Jos\'e P. Palao}
\affiliation{Departamento de F\'isica Fundamental II and IUdEA, 
Universidad de La Laguna, Spain,
La Laguna 38204, Spain}

\author{Christiane P. Koch}
\affiliation{Theoretische Physik, Universit\"at Kassel,
  Heinrich-Plett-Str. 40, 34132 Kassel, Germany}

\date{\today}

\begin{abstract}
  Shaped pulses obtained by optimal control theory often possess
  unphysically broad spectra. In principle, the spectral width of a
  pulse can be restricted by an additional constraint in the
  optimization functional. However, 
  it has so far been impossible  to impose 
  spectral constraints while strictly guaranteeing monotonic
  convergence. 
  Here, we show that Krotov's method allows for simultaneously
  imposing temporal and spectral constraints without perturbing
  monotonic convergence,   provided the constraints can be expressed as 
  positive semi-definite quadratic forms. 
  The optimized field is given by an integral equation which can
  be solved efficiently using the method of degenerate kernels. We
  demonstrate that Gaussian filters suppress undesired frequency
  components in the control of non-resonant two-photon absorption.
\end{abstract}

\maketitle

\section{Introduction}
\label{sec:intro}

Optimal control theory (OCT) is a versatile mathematical tool to find
external fields that drive the dynamics of
a quantum system toward a desired outcome~\cite{RiceBook}. 
The controls are, e.g., the electric field of a laser pulse or the
magnetic field amplitude of radio-frequency (RF) pulses. The underlying 
mechanism enabling the control are quantum interferences of light and
matter~\cite{RiceBook,ShapiroBook}. 
OCT consists in formulating the physical target as a functional of the
field which is 
then optimized. Typically, many solutions to the control problem
exist~\cite{RabitzScience04}, and it depends on additional constraints
which of these solutions is found by an OCT algorithm. Such additional
costs can be used to identify solutions that are feasible 
in control experiments, for example in feedback loops with
shaped femtosecond laser pulses~\cite{RabitzScience00} or sequences of
RF pulses in high-resolution nuclear magnetic
resonance~\cite{SkinnerJMR03}. The 
constraints ensure, for example, a maximally allowed amplitude or smoothly
switching the pulses on and off~\cite{SundermannJCP99}. 
In principle, a constraint to ensure a given spectral
width of the pulse can be formulated analogously~\cite{KochPRA04}.  
It is highly desirable to include such a constraint 
since the spectral width is fixed in a given
experiment. In order to compare theoretically calculated and
experimentally obtained pulses, it is  necessary to restrict the
bandwidth of the calculated pulses to the experimental value. 
However, so far it has  been impossible to impose spectral constraints
while strictly guaranteeing monotonic convergence of the optimization
algorithm. Without a spectral constraint, the optimized pulses 
often possess extremely broad spectra with frequency components that
are physically not necessary and cannot be realised experimentally,
see, e.g., Ref.~\cite{KochPRA04}. 

To obtain control over the frequency components of the optimal pulse, 
two alternatives to imposing spectral constraints as part of the
optimization functional have recently been discussed: 
(i) The field can be expanded into frequency components,
and the expansion coefficients, not the field itself, are
optimized~\cite{SkinnerJMR10}. This approach requires a
concurrent update of the field $\epsilon(t_i)$ for all $t_i$ at once and cannot be
combined with a sequential update. (ii) The optimized field can be
filtered at the end of each iteration step to
eliminate undesired frequency
components~\cite{WerschnikJOB05,GollubPRL08,LapertPRA09,SchroederNJP09,GollubPCCP10,MotzoiPRA11,IdoPRA12}.   
The challenge consists in
implementing the filtering in a way that does not destroy convergence
of the algorithm. Formally, a filter can be obtained from a cost
functional. However, the corresponding Lagrange multiplier which is
decisive for the 
convergence of the algorithm, remains
indetermined~\cite{GollubPRL08,GollubPCCP10}. An educated guess for
the Lagrange multiplier was shown to work under certain assumptions on
the pulse and for sufficiently slow increase of the undesired
frequency components~\cite{GollubPRL08,GollubPCCP10}. 
It is nonetheless dissatisfying that monotonic convergence cannot be
ensured in general. An alternative filtering approach that strictly
enforces convergence interpolates 
between the unfiltered field obeying monotonic convergence and the
completely filtered field destroying convergence. The strength of the
filter is then chosen in such a way that the filter barely avoids
breaking the convergence~\cite{LapertPRA09}. This approach comes with
considerable extra numerical effort since the interpolation
requires additional optimization runs for each value of the
interpolation parameter.  

Here, we demonstrate that spectral constraints can be included in the
optimization functional without perturbing monotonic convergence 
using Krotov's method \cite{Konnov99,SklarzPRA02,JosePRA03,ReichKochJCP12}. 
The spectral constraint is expressed via its Fourier transform as an
integral over time. The corresponding integral kernel  must be written
as a positive semi-definite quadratic form. We show that this is the 
only requirement that needs to be met to ensure monotonic
convergence. The modified update formula for the field corresponds to 
a Fredholm integral equation of the second kind which can be solved
efficiently using the method of degenerate kernels. We apply Krotov's
method including spectral constraints to the optimal control of
non-resonant two-photon absorption.

\section{Spectral constraints in Krotov's method}
\label{sec:method}

In optimal control theory, 
the optimization problem is formulated by stating the 
target and additional constraints in functional form, 
\begin{equation}
  \label{eq:J}
  J[\{\psi_k\}, {\epsilon}] = J_T[\{\psi_k(T)\}]
  +J_a[\epsilon]
  +J_b[\{\psi_k\}] \,,  
\end{equation}
where $J_T$ denotes the target at final time $T$ and 
$\{\psi_k(t)\}$ a set of state vectors describing the time evolution of the
system. $\epsilon(t)$ is a real function representing the control
variable, e.g., the electric field amplitude of a laser pulse.
All additional constraints are assumed to depend either on the control
or on the states, 
\begin{eqnarray}\label{eq:Jab}
  J_a =\int_0^T{g_{a}(\epsilon,t) ~ dt } \,, \quad
  J_b =\int_0^T{g_{b}(\{\psi_k\},t) ~ dt }. 
\end{eqnarray}
A common choice for $g_a(\epsilon,t)$ minimizes the pulse intensity or
change in pulse intensity \cite{JosePRA03},
\begin{equation}
  \label{eq:amplitude constraint}
  g_{a}(\epsilon,t)=\frac{\lambda_0}{S(t)}
  \left[{\epsilon(t) - \epsilon^{(0)}(t)}\right]^2 
  = \frac{\lambda_0}{S(t)}\left[\Delta \epsilon(t)\right]^2 \,,
\end{equation}
with $\lambda_0$ a weight to favor solutions with lower pulse
amplitude  and $S(t)$ a shape function to
smoothly switch the pulse on and off. $J_b$ can be used to restrict
the time evolution to a subspace of the Hilbert space or to optimize a
time-dependent target, see Ref. \cite{ReichKochJCP12} and references
therein. 

Minimization of the functional~\eqref{eq:J} yields a set of coupled
equations for the states and the control. 
The non-linear optimization method developed by Konnov and
Krotov~\cite{Konnov99} provides  
a general, monotonically convergent algorithm. Given
Eq.~\eqref{eq:amplitude constraint} for $g_a$, it updates the control 
at iteration step $i+1$ according to~\cite{ReichKochJCP12}
\begin{equation}
  \label{eq:update}
  \epsilon^{(i+1)}(t) = \epsilon^{(i)}(t) + \frac{S(t)}{\lambda_0}
  \mathfrak{Im} \left\{\sum_k \Braket{\chi_k^{(i)}(t)|
      \frac{\partial \Op H}{\partial \epsilon}|\psi_k^{(i+1)}(t)}  
    + \frac{1}{2} \sigma(t) \sum_k \Braket{\Delta\psi_k(t)|
      \frac{\partial \Op H}{\partial\epsilon}|\psi^{(i+1)}_k(t)} \right\} \,,
\end{equation}
where 
$\Ket{\Delta\psi_k(t)}=|\psi^{(i+1)}_k(t)\rangle-|\psi^{(i)}_k(t)\rangle$
and $\Op H$ the Hamiltonian of the system. The 
adjoint states $\Ket{\chi_k(t)}$ are propagated backwards in time with
the boundary condition $\Ket{\chi_k(T)}$ determined by the final-time
target $J_T$. The choice of
the function $\sigma(t)$ allows for ensuring 
monotonic convergence~\cite{Konnov99}. The specific form of 
$\sigma(t)$ depends on the optimization functional and the equations
of motion. It can be estimated analytically or determined numerically,
based on the optimization history~\cite{ReichKochJCP12}.  

Constraints on the spectrum of the control have to be included in
the cost functional $J_a$. Monotonic convergence requires a
well-defined sign of $J_a$~\cite{JosePRA03,ReichKochJCP12}. A general
expression that fulfills this requirement is obtained by writing $J_a$
as a quadratic form. In frequency domain, necessary to formulate spectral
constraints, the cost functional thus becomes 
\begin{eqnarray}
  \label{eq:Jw}
  J_{a}(\epsilon) =
  \int_{-\infty}^{\infty}{\Delta\epsilon(\omega) \bar{K}(\omega)
    \Delta\epsilon^{*}(\omega) ~ d\omega}  
  = \frac{1}{2\pi}
  \int_{-\infty}^{\infty}{\int_{-\infty}^{\infty}
    {\Delta\epsilon(t)K(t-t')\Delta\epsilon(t')~ dt'} ~ dt} \,, \nonumber
\end{eqnarray}
where a real kernel function $\bar{K}$ in frequency domain
and its Fourier transform $K$ in time domain have been introduced. 
The desired spectral constraints are thus implemented by the kernel
function. Given Eq.~\eqref{eq:Jw} for $J_a$,  
the function  $g_{a}$, defined in Eq.~\eqref{eq:Jab}, takes the form 
\begin{equation}
  \label{eq:frequency constraint}
  g_a(\epsilon,t) = \frac{1}{2\pi} \int_{0}^{T}
  {\Delta\epsilon(t) K(t-t') \Delta\epsilon(t') ~ dt'} \,.
\end{equation}
Since the field and thus the change in the field are zero outside of
the interval $[0,T]$, integration can be restricted to $[0,T]$.
In Krotov's method, monotonic convergence can be ensured if the kernel
$K(t-t')$ is positive semi-definite~\cite{ReichKochJCP12}.
This follows directly from the condition for the change of the
functional due to changes in the control to be 
positive \cite{JosePRA03,ReichKochJCP12} which in turn translates into $g_a$
being a convex function. 
Equivalently in frequency domain, $\bar{K}(\omega)$ has to be 
positive semi-definite.
Monotonic convergence is therefore guaranteed if 
\begin{equation}
  \label{eq:Kcondition}
\bar{K}(\omega)\geq0 \quad \forall ~ \omega\,.   
\end{equation}

Since derivation of the update equation requires
evaluation of $\frac{\partial g_a}{\partial \epsilon}$ as a function of
time \cite{SklarzPRA02,JosePRA03,ReichKochJCP12}, 
the  Fourier transform of $\bar{K}(\omega)$ should have a closed
form in addition to being positive semi-definite. For numerical
stability, it is furthermore desirable to use smooth kernels. A
suitable choice fulfilling these requirements are Gaussian kernels,
\begin{eqnarray}
  \bar{K}(\omega) &=& \lambda_a - \sum_i \frac{\lambda_b^i}{2} 
    \left[ e^{-\frac{(\omega-\omega_i)^2}{2\sigma_i^2}} + 
      e^{-\frac{(\omega+\omega_i)^2}{2\sigma_i^2}} \right] \,,\nonumber\\
    \label{eq:kernel}
    K(t-t') &=&  2 \pi \lambda_a \delta(t-t') 
    - \sum_i \lambda_b^i \sqrt{2\pi \sigma_i^2} 
    \cos[\omega_i (t-t')] e^{-\frac{\sigma_i^2 (t-t')^2}{2}} \,.
\end{eqnarray}
Note that we choose symmetric Gaussian kernels since we consider here
real fields. An extension to complex controls is straightforward.
For (approximately) non-overlapping Gaussians in frequency domain,
monotonic convergence is obtained  if 
\begin{equation}
  \label{eq:condlambda}
  \lambda_b^i \leq 2 \lambda_a ~ \forall\,i  \,.
\end{equation}
The first term in Eq.~\eqref{eq:kernel}
reproduces Eq.~\eqref{eq:amplitude constraint} with 
$\lambda_0=\lambda_a$ and $S(t)=1$. 
For $\lambda_b^i>0$, the kernel~\eqref{eq:kernel} implements a
frequency pass for $\Delta\epsilon(t)$
around the frequencies $\omega_i$. 
For $\lambda_b^i<0$, a frequency filter for $\Delta\epsilon(t)$ 
around the frequencies $\omega_i$ is obtained. 
Due to the condition~\eqref{eq:condlambda}, frequency passes are not
guaranteed to be effective, i.e., the $\lambda_b^i$ might be too small
for the spectral constraint to gain sufficient weight. 
For frequency filters, no such restriction exists. A work-around to
create effective frequency passes consists therefore in
 adding up sufficiently many
frequency filters. Moreover, an amplitude constraint
with non-constant shape function can be reintroduced additively in
time domain for $\lambda_b^i < 0$, setting $\lambda_a =0$. This does
not perturb monotonic convergence since both amplitude and frequency
constraint preserve monotonic convergence individually.  

Following the prescription of Ref.~\cite{ReichKochJCP12}, the 
update equation for Gaussian band filters around frequencies
$\omega_i$ and an additional amplitude constraint imposed by a shape
function $\lambda_0 / S(t)$ is obtained as  
\begin{eqnarray}
  {\epsilon^{(i+1)}(t)} & = & \epsilon^{(i)}(t) + \sum_i
  \frac{\lambda_b^i S(t)}{2 \pi \lambda_0} \sqrt{2\pi \sigma_i^2}
  \int_0^T{\cos[\omega_i (t-t')]\, e^{-\frac{\sigma_i^2
        (t-t')^2}{2}} \left({\epsilon^{(i+1)}(t')} -
      \epsilon^{(i)}(t')\right) ~ dt'} \nonumber \\ 
  & & + \frac{S(t)}{\lambda_0} \mathfrak{Im} \left\{\sum_k
    \Braket{\chi_k^{(i)}(t)|\frac{\partial \Op H}{\partial
        \epsilon}|\psi_k^{(i+1)}(t)}  
    + \frac{1}{2} \sigma(t) \sum_k
    \Braket{\Delta\psi_k(t)|\frac{\partial \Op H}{\partial
        \epsilon}|\psi^{(i+1)}_k(t)} \right\} .\ 
  \label{eq:newupdate}
\end{eqnarray} 
This is an implicit equation for $\epsilon^{(i+1)}(t)$. It is possible
to rewrite Eq.~\eqref{eq:newupdate} as a Fredholm integral equation of
the second kind for $\Delta \epsilon(t) =
\epsilon^{(i+1)}(t) - \epsilon^{(i)}(t)$, 
\begin{equation}
  \label{eq:Fredholm}
  \Delta \epsilon(t) = I(t) + \gamma \int_0^T{\mathcal{K}(t,t') \Delta
    \epsilon(t') ~ dt'} .\ 
\end{equation}
The inhomogeneity $I(t)$ depends on the unknown states 
$\{\psi_k^{(i+1)}(t)\}$. They can be approximated
by calculating $\Delta\epsilon(t)$ according to
Eq.~\eqref{eq:update}, i.e., without frequency constraints. 
Propagating the states under that field yields an approximation of
$I(t)$. In our applications this turned out to be sufficient. However,
if the quality of the resulting approximation of $I(t)$ is not good
enough, the field obtained from a first solution of the Fredholm
equation can be used to propagate the states and obtain an improved
approximation of $I(t)$. This procedure 
can be repeated iteratively  until the desired accuracy is reached. 
The remaining question is then how to solve the integral
equation~\eqref{eq:Fredholm}. 

Often, Fredholm equations of the second kind are solved
numerically~\cite{EncycMath} by quadrature of the integral,  
\[
 \int_0^T \mathcal{K}(t,t') \Delta\epsilon(t') ~ dt' \simeq
   \sum_{j=1}^N w_j \mathcal{K}(t,t_j) \Delta\epsilon(t_j)
\]
such that
\[
\Delta\epsilon(t_k) \simeq 
I(t_k) + \gamma  \sum_{j=1}^N w_j \mathcal{K}(t_k,t_j) \Delta\epsilon(t_j)\,,
\]
or collocation, i.e., expanding $\Delta\epsilon(t)$ into orthonormal
basis functions $c_j(t)$ on $[0,T]$,
\[
\Delta\epsilon(t) = \sum_{j=1}^N a_j c_j(t)\,.
\]
In both cases, solution of the integral equation is
reduced to solving a system of linear equations. 
Alternatively, a Fredholm equation of the second kind can be solved by 
approximating $\mathcal{K}(t,t')$ by a degenerate
kernel, $\mathcal{K}_N(t,t')=\sum_{j=1}^N\alpha_j(t)\delta_j(t')$
\cite{EncycMath}. Solution of a Fredholm degenerate integral
equation again reduces to solving a system of linear equations.
For our purposes, an approach based  on degenerate
kernels~\cite{Volk79,Volk85} turns out to be the best option.
It is more stable than collocation and 
similar to the quadrature of the integral but more direct since 
the kernel rather than the integral is approximated.
The solution to Eq.~\eqref{eq:Fredholm} is then given by
\begin{equation}
  \label{eq:solFred}
  \Delta\epsilon(t) = I(t) + \sum_{j=0}^N X_j \alpha_j(t)
\end{equation}
with $\alpha_j(t)$ defined in Eq.~\eqref{eq:alpha} and
$X_j$ the solution of the system of linear
equations~\eqref{eq:lin}. 

\section{Control of non-resonant two-photon absorption}
\label{sec:TPA}

We apply Krotov's method including spectral constraints,
Eq.~\eqref{eq:newupdate}, to non-resonant two-photon absorption
in sodium atoms.
The goal is to transfer population from level $\ket{3s}$ to 
$\ket{4s}$. Due to selection rules, this is possible only by
absorption of two photons with the 
transition dipoles provided by the off-resonant $\ket{np}$ levels with
the main contribution coming from $\ket{3p}$.
We do not invoke an adiabatic elimination of all off-resonant levels,
i.e., our Hamiltonian includes
$\{\ket{3s},\ket{4s},\ket{np}\}$ with $n=3,\ldots,8$ and the corresponding
$s-p$ transition dipole moments, taken from Ref.~\cite{NIST}.
Within this model, two control strategies are available to transfer
population from $\ket{3s}$ to $\ket{4s}$ -- resonant two-color
one-photon transitions with frequencies $\omega_{3s,3p}$ and
$\omega_{3p,4s}$  or an off-resonant two-photon transition with
frequency close to $\omega_{3s,4s}/2$.

Non-resonant two-photon absorption has been studied experimentally for
$ns$ to $(n+1)s$ transitions in alkali atoms in the
weak~\cite{MeshulachNature98,MeshulachPRA99,PraekeltPRA04},
strong~\cite{TralleroPRA05,TralleroPRL06,TralleroPRA07} and intermediate field 
regime~\cite{ChuntonovPRA08,ChuntonovJPB08,ZoharPRL08,ChuntonovPRA10}. 
To date, optimal control calculations of non-resonant two-photon
absorption have been hampered by a spectral spread of the field.
The resulting spectral widths by far exceed
experimentally realistic values.
As a result, only solutions using one-photon transitions are
found while the experimental result of \textit{non-resonant} two-photon
control~\cite{MeshulachNature98,MeshulachPRA99,PraekeltPRA04,TralleroPRA05,TralleroPRL06,TralleroPRA07,ChuntonovPRA08,ChuntonovJPB08,ZoharPRL08,ChuntonovPRA10} 
could not be reproduced. 
Here we employ optimal control theory with spectral constraints 
to enforce a non-resonant two-photon solution. We use Gaussian
frequency filters around the 
one-photon frequencies to suppress
resonant dipole transitions. 

\begin{figure}[tb]
  \centering
  \includegraphics[width=0.75\linewidth]{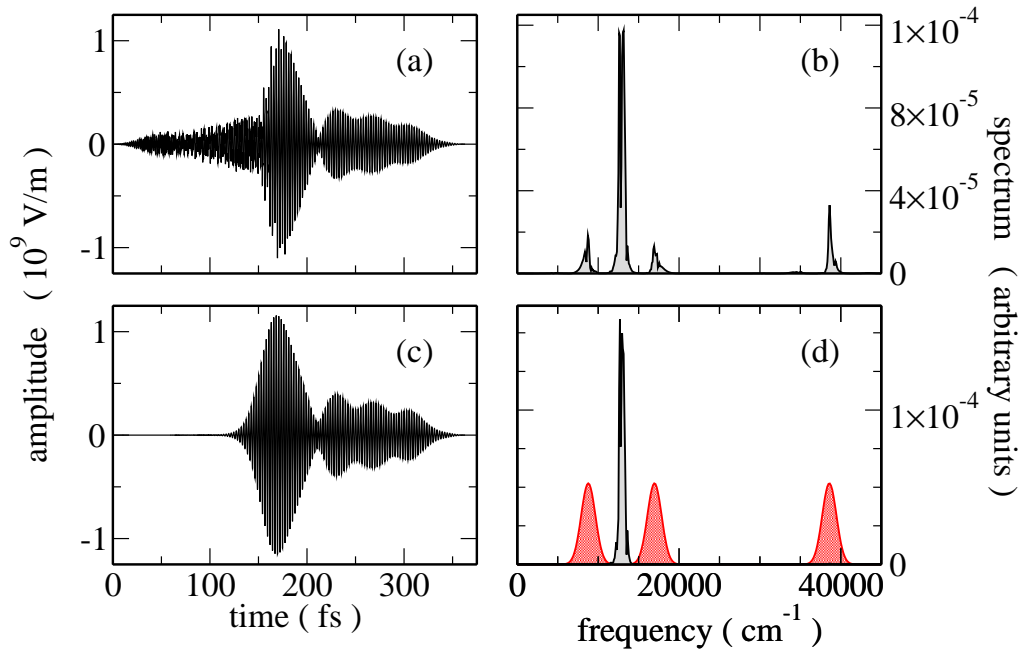}
  \caption{
    Optimized pulses and their spectra with (c+d) and without (a+b)
    spectral constraint. The Gaussian filters employed in the spectral
    constraint are shown in red. 
  }
  \label{fig:TPA}
\end{figure}
\begin{figure}[tb]
  \centering
  \includegraphics[width=0.75\linewidth]{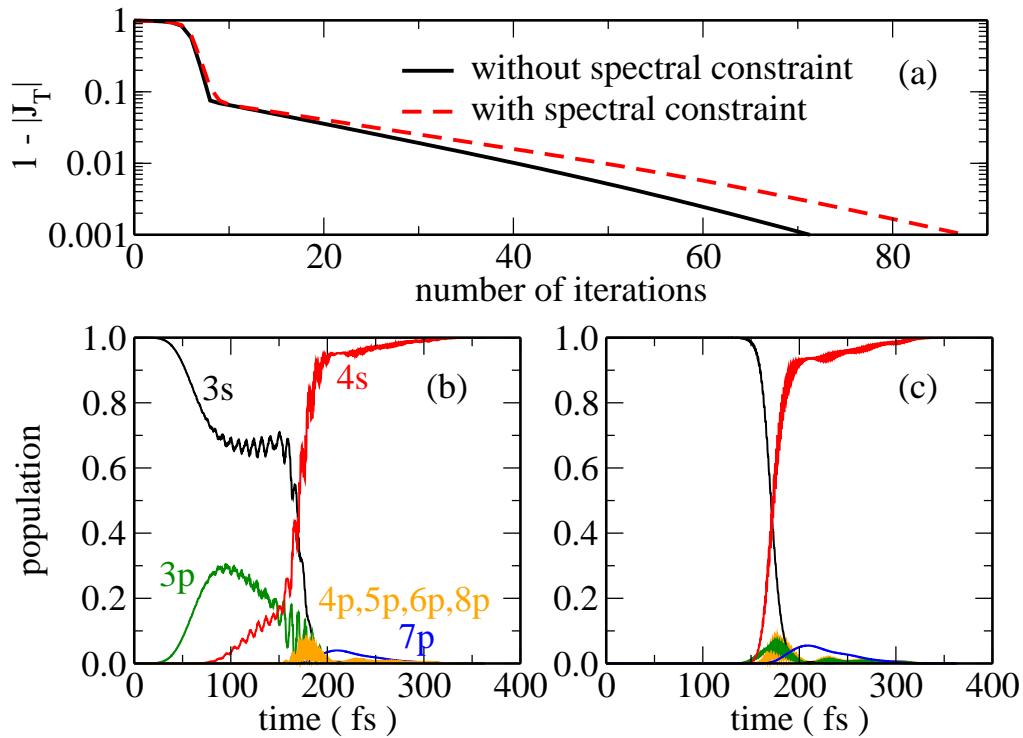}  
  \caption{
    Convergence toward the optimum (a) and dynamics under the
    optimized pulses with (c) and without (b) spectral constraint.} 
  \label{fig:conv_TPA}
\end{figure}

Figure~\ref{fig:TPA}  compares the optimal pulses and their spectra
obtained by Krotov's method with (bottom panel) and without (top
panel) spectral constraint, cf. Eqs.~\eqref{eq:newupdate} and
\eqref{eq:update}. 
The frequency filters around the one-photon transition frequencies are
indicated in red in Fig.~\ref{fig:TPA}(d). 
The central frequency of the guess pulse is taken to be exactly the
two-photon transition frequency. Its peak amplitude is about a fourth
of that of a two-photon $\pi$-pulse. Despite the guess
pulse being fairly close to a non-resonant two-photon solution, the
optimization algorithm yields a pulse that uses the resonant
one-photon transitions, cf. the three small peaks 
in Fig.~\ref{fig:TPA}(b). The use of one-photon transitions is also
reflected in the dynamics under the optimized pulse which shows a
significant population of the $|3p\rangle$ level, cf. the green line
in  Fig.~\ref{fig:conv_TPA}(b). It is rationalized in terms of the 
intensity which should increase as little as possible according to the
constraint~\eqref{eq:amplitude constraint} and resonant transitions
requiring a lot less intensity than non-resonant ones. 
Increasing the spectral width comes at no 'cost' for the optimization
algorithm when no spectral constraint is present. Thus solutions that
use resonant one-photon transitions and have a broad spectral width
are the natural ones for optimization without spectral constraint. 
Once the spectral constraint is included, the optimization algorithm
increases the pulse amplitude until a two-photon Rabi frequency of
$\pi$ is hit. The spectrum of the optimal
pulse is hardly modified compared to that of the guess pulse.

Imposing an additional constraint results in a more difficult
optimization problem. This is illustrated by
Fig.~\ref{fig:conv_TPA}(a) 
which compares the convergence toward the optimum for optimization
with and without the spectral constraint. In order to reach the
optimum within an 'error', $\varepsilon=1-|J_T|$, of $10^{-3}$
the number of iterations is increased from 71 to 87.
The slower convergence of the algorithm with
spectral constraint is attributed to optimization under two
conflicting costs -- keeping the intensity as low as possible while
avoiding certain spectral regions. The algorithm needs to
balance the two conflicting costs, which results in a more difficult
optimization problem. 

While the increase in the number of iterations, when adding
the spectral constraint, is comparatively moderate, a CPU time of about
370$\,$s is needed for 10 iterations, compared to only 6$\,$s for the
algorithm without spectral constraint. This is due to the additional
numerical effort required in order to solve the Fredholm equation. 
This effort scales with the number of time grid points but is
independent of the complexity of the system. The comparison
of the CPU time required with and without the spectral constraint will
be much more favorable for more complex systems. Then most of the
CPU time will be spent for the time propagation whereas the
solution of the Fredholm equation represents a comparatively small
add-on. Moreover, the numerical effort for solving the Fredholm
equation can be further reduced by exploiting the bandedness of the
matrix in Eq.~\eqref{eq:lin}. 

\section{Summary}
\label{sec:concl}

We have derived an extension of Krotov's method for quantum optimal
control that allows for including constraints on
the control in frequency and time domain at the same time.
The key is to ensure a well-defined sign of the integral over the
constraint which we have achieved by expressing the constraint as a
quadratic form. Gaussian kernels, to be used either as frequency
passes or as frequency filters, turn out to be the most practical
choice. Frequency passes may be inefficient due to a limit on the
weight of the constraint, whereas frequency filters can be employed
without restriction. 

The update equation that we obtain for Gaussian frequency filters
is an implicit equation in the control which takes the form of a
Fredholm integral equation of the second kind. It can be solved 
accurately and efficiently using the method of degenerate
kernels~\cite{Volk79,Volk85}. Our results for non-resonant two-photon
absorption in sodium atoms show an excellent restriction on the
spectrum of the optimized pulse. The new algorithm thus allows for 
reproducing experimentally known control strategies for strong-field
non-resonant two-photon
absorption~\cite{TralleroPRA05,TralleroPRL06,TralleroPRA07}.  
It can also be used in conjunction with quasi-Newton methods in order
to achieve faster convergence~\cite{EitanPRA11}. In future work, we
will discuss in detail how the spectral constraint allows for steering the
optimization pathway in the control landscape~\cite{Jose13}. 

\section*{Acknowledgments}
We would like to thank Ronnie Kosloff for many valuable
discussions.  Financial support from the Deutsche
Forschungsgemeinschaft (grant No. KO 2301/2) 
and by the Spanish MICINN (grant No. FIS2010-19998)
is gratefully acknowledged.

\appendix

\section{Method of degenerate kernels for the 
  numerical solution of  
  Fredholm equations of the second kind}

To simplify notation, we map the time interval from $[0,T]$ to
$[0,1]$. 
A degenerate kernel is obtained by a tensor product ansatz for the
true kernel,  
\begin{equation*}
  \mathcal{K}(t,t') \simeq \sum_{j,k=0}^N d_{jk} \alpha_j(t) \beta_k(t')\,,
\end{equation*}
with $\delta_j(t') = \sum_{k=0}^N d_{jk} \beta_k(t')$, taking 
the basis functions to be~\cite{Volk79,Volk85}
\begin{equation}\label{eq:alpha}
  \alpha_{j}\left(t\right)=\beta_{j}\left(t\right)=
\begin{cases}
  1-N\left|t-\frac{j}{N}\right|, & \frac{j-1}{N}\leq t\leq\frac{j+1}{N} \\
  0, & \text{else} \end{cases} \,.
\end{equation}
$N$ is the order of the approximation. 
At the grid points $t=\frac{u}{N}$, $t'=\frac{v}{N}$,
\begin{eqnarray*}
  \mathcal{K}_N\left(\frac{u}{N}, \frac{v}{N}\right) =
  \sum_{j,k=0}^N d_{jk} \delta_{ju} \delta_{kv}\,.
\end{eqnarray*}
The choice of basis functions suggests for the coefficients  
\[
d_{jk} = \mathcal{K}_N\left(\frac{j}{N},\frac{k}{N}\right)
= \mathcal{K}_N(t_j,t_k) \,,
\]
such that $\mathcal{K}_N$ reasonably approximates $\mathcal{K}(t,t')$
on a time grid of size $N+1$.   

It can be shown 
that the solution to Eq.~\eqref{eq:Fredholm} is given
by Eq.~\eqref{eq:solFred} with $X_j$ the solution of the following
system of linear equations,
\begin{equation}
  \label{eq:lin}  
\left[\openone_{N+1} -\gamma\boldsymbol{\mathsf{C}} \right] 
\vec{X} = \gamma \vec{b} \,,
\end{equation}
with matrix elements
\[
C_{jk} = \sum_{i=0}^{n}K\left(t_{j},t_{i}\right)\int_{0}^{1}\alpha_{i}\left(t\right)\alpha_{k}\left(t\right)\ dt\equiv\sum_{i=0}^{n}K\left(t_{j},t_{i}\right)A_{ik} \,,
\]
where 
\[
A_{ik} = \int_{0}^{1}\alpha_{i}\left(t\right)\alpha_{k}\left(t\right)\ dt=\begin{cases}
\frac{1}{3n}, & \text{for }i=k=0\;\text{or}\;i=k=n\\
\frac{2}{3n}, & \text{for }i=k,1\leq i\leq n\\
\frac{1}{6n}, & \text{for }i=k+1\text{ or }i=k-1\\
0, & \text{else}\end{cases}
\]
and 
\[
b_{k} = \int_{0}^{1}I\left(t\right)\left[\sum_{i=0}^{n}K\left(t_{k},t_{i}\right)\alpha_{i}\left(t\right)\right]\ dt  \,.
\]


\end{document}